\RequirePackage{fix-cm}
\documentclass[smallextended]{svjour3}       
\smartqed  
\usepackage{graphicx}
\begin{document}

\title{Gravitational wave science with laser interferometers and pulsar timing 
}


\author{Alberto Sesana  
}


\institute{Max-Planck-Institut f{\"u}r Gravitationsphysik, Albert Einstein Institut, Am M\"uhlenberg 1, 14476 Golm, Germany \\
              \email{alberto.sesana@aei.mpg.de}           
}

\date{Received: date / Accepted: date}

\maketitle

\begin{abstract}
Within this decade gravitational wave detection will open a new observational window on the Universe. Advanced ground based interferometers covering the kHz frequency range will be online by 2016, and it is foreseeable the announcement of a first detection within five years. At the same time, a worldwide effort of detecting low frequency waves (in the nHz regime) by timing ultra-precise millisecond pulsars is rapidly growing, possibly leading to a positive detection within this decade. The mHz regime, bridging these two windows, is the realm of space based interferometers, which might be launched in the late 20s. I provide here a short overview of the scientific payouts of gravitational wave astronomy, focusing the discussion on the low frequency regime (pulsar timing and space based interferometry). A detailed discussion of advanced ground based interferometer can be found in Patrick Brady's contribution to this proceedings series.

\end{abstract}

\section{Introduction}
\label{intro}

Our current knowledge of the Universe comes primarily from the observation of photons released in the physical processes driving the formation and evolution of stars and galaxies. However, black holes (BHs, among the most exciting astrophysical objects) are 'dark' in nature, and can be probed in a clean way only by {\it directly} capturing dynamical modification of the spacetime metric induced by their accelerated motion and radiated away at the speed of light. Those are better known as gravitational waves (GWs). GWs are not exclusively produced by BHs, but by any accelerated mass with a time varying mass quadrupole moment, however they are so weak, that the only potentially observable sources involve astrophysical compact objects. Among the most appealing sources are: catastrophic events like supernovae (SNe) explosions and gamma ray bursts (GRBs) leaving behind a compact remnant, fast rotating neutron stars (NSs) with some irregularity in their mass distribution, and, in particular, compact binaries. The latter class of sources include any combination of stellar compact objects (white dwarfs (WDs), NSs and BHs) and systems involving massive and supermassive BHs (to which we will generally refer as MBHs in the following): extreme mass ratio inspirals (EMRIs) of stellar compact objects into a MBH, and MBH binaries.  

The signals emitted by the aforementioned sources have so far silently been lost in space, however, within this decade the detection of GWs may indeed be a reality, opening a completely new window on the Universe. While signals coming from compact stars and binaries fall in the observational domain of operating and planned ground based interferometers (such as the advanced LIGO \cite{harry10}, VIRGO \cite{acernese06}, KAGRA \cite{somiya12}, and the proposed 'third generation' Einstein Telescope (ET) \cite{punturo10}), MBH binaries and EMRIs are expected to be among the primary actors on the upcoming low frequency stage, where the $10^{-4}-10^{-1}$Hz window is going to be probed by spaceborne interferometers like the Laser Interferometer Space Antenna (LISA) \cite{bender98}, or the recently proposed European development eLISA/NGO \cite{amaro12a,amaro12b}. At even lower frequencies, precision timing of an array of millisecond pulsars (i.e. a pulsar timing array, PTA) provides a unique opportunity to get the very first low-frequency detection. The European Pulsar Timing Array (EPTA) \cite{ferdman10}, the Parkes Pulsar Timing Array (PPTA) \cite{man12} and the North American Nanohertz Observatory for Gravitational Waves (NANOGrav) \cite{jenet09}, joining together in the International Pulsar Timing Array (IPTA) \cite{hobbs10}, are constantly improving their sensitivity in the frequency range of $\sim10^{-9}-10^{-6}$ Hz, and the planned Square Kilometer Array (SKA) \cite{aha13} will provide a major leap in sensitivity.  

In this contribution I will focus on the most relevant GW {\it astrophysical sources}, meaning that I will not touch the possibility of detecting {\it cosmological} backgrounds from e.g., inflation, phase transitions and defects in the early Universe and cosmological strings. An overview of the latter can be found in \cite{binetruy13}. Among all astrophysical sources particular attention will be devoted to binaries of all kind, which are the one we understand better and that will probably carry the most valuable and cleanest astrophysical information. Excellent extensive reviews on the subject can be found in \cite{thorne95,hughes03}. After a short general introduction about the principles of GW generation and detection (Section \ref{gws}), I will only briefly touch the ground based interferometer science in Section 3. Section 4 is devoted to space-based interferometers and Section 5 to PTAs.  

\section{Gravitational waves: generation and detection principles}
\label{gws}

\subsection{Gravitational wave generation}
The existence of GWs was one of the first predictions of Einstein's General Relativity (GR) \cite{einstein16}. However, universal acceptance of their existence came only in the '80s, thanks to the excellent agreement between theoretical predictions \cite{damour83} and observations of the Hulse-Taylor binary pulsar \cite{taylor89}. GWs are ripples in the fabric of the spacetime propagating at the speed of light. Accelerating masses generate GWs pretty much in the same way as accelerating charges generate electromagnetic waves. By expanding the mass distribution of the source into multipoles, conservation laws enforce GWs coming from the mass monopole and mass dipole to be identically zero, so that the first contribution to GW generation comes from the mass quadrupole moment $Q$. The GW amplitude is therefore proportional to the second time derivative (acceleration) of $Q$. Moreover, energy conservation enforces the amplitude to decay as the inverse of the distance to the source, $D$. A straightforward dimensional analysis shows that the amplitude of a GW is of the order of \cite{hughes03}
\begin{equation}
h=\frac{G}{c^4}\frac{1}{D}\frac{d^2Q}{dt^2}.
\label{eqstrain}
\end{equation}
In order to generate GWs we therefore need accelerating masses with a time varying mass-quadrupole moment. The prefactor $G/c^4$ implies that these waves are {\it tiny}, so that the only detectable effect is produced by massive compact astrophysical objects, as enumerated in the Introduction. A detailed derivation of the GW properties is beyond the scope of this review, I shell just mention that GWs are transverse ,i.e. they act in a plane perpendicular to the wave propagation, and (at least in GR) have two distinct polarizations, usually referred to as $h_+$ and $h_\times$ (see cartoon in \cite{thorne95}). Their quadrupolar nature implies that perpendicular directions experience 'opposite' squeezing, so that if the GW propagates in the $z$ direction, the metric shrinks in the $x$ direction while expanding in the $y$ direction and viceversa in a oscillatory fashion.

Using equation (\ref{eqstrain}) we can infer approximate GW amplitudes for typical sources. Consider, for simplicity, a circular BH binary system characterized by a Keplerian frequency $f_K=f/2${\footnote{A circular binary emits GWs at a frequency which is twice the Keplerian binary frequency, whereas an eccentric one emits a more complicated wave covering the spectral range $nf_k$, where $n$ is an integer index.}}, chirp mass ${\cal M}=M_1^{3/5}M_2^{3/5}/(M_1+M_2)^{1/5}$ ($M_1>M_2$ are the individual masses of the two BHs), and distance to the observer D, equation (\ref{eqstrain}) results in a typical strain of 
\begin{equation}
h\approx 10^{-22}\left(\frac{D}{100{\rm Mpc}}\right)^{-1}\left(\frac{{\cal M}}{{\rm M_{\odot}}}\right)^{5/3}\left(\frac{f}{100{\rm Hz}}\right)^{2/3}.
\label{strainbin}
\end{equation}
The maximum emission frequency is approximately given by twice the Keplerian frequency of the binary at the innermost stable circular orbit (ISCO, binary separation $a=6GM_1/c^2=3R_S$, where $R_S$ is the Schwarzschild radius of the more massive object in the binary)
\begin{equation}
f_{\rm ISCO}\approx10^{4}{\rm Hz} \frac{{M_1}}{{\rm M_{\odot}}}.
\label{fisco}
\end{equation}
Equations (\ref{strainbin}) and (\ref{fisco}) are indeed valid for any binary of compact objects, with one important difference: when the binary involves 'extended objects' (like NSs or WDs), the maximum frequency is given by the 'contact' separation of the two bodies. For example, the maximum WD-WD binary frequency is roughly $0.1 Hz$, beyond which, the two body merge in a single object. 

Plugging in the mass quadrupole moment for a rotating ellipsoid with ellipticity $\epsilon$ into equation (\ref{eqstrain}) gives instead 
\begin{equation}
h\approx 10^{-26}\left(\frac{D}{1{\rm kpc}}\right)^{-1}\left(\frac{f}{1000{\rm Hz}}\right)^{2}\left(\frac{\epsilon}{{10^{-8}}}\right)\left(\frac{Q}{{10^{45}{\rm g m}^2}}\right),
\label{strainspin}
\end{equation}
where we normalized to typical values of a rapidly spinning NS. On the other hand, predicting the waves coming from collapses and explosions is much more difficult and requires detailed 3D simulations of the process. A fair bet is to assume that the explosions involve the non-axisymmetric acceleration of some substantial fraction of $Q$, to a speed which is a fraction of $c$. This automatically sets the range of amplitudes and frequencies of interest to $h\sim 10^{-23}-10^{-20}$ and $f\sim 10^{2}-10^{3}$Hz respectively (see \cite{ott09} for an excellent review).   

\subsection{Gravitational wave detection}
The quantity $h$ given in equation {\ref{eqstrain}} is usually referred to as 'strain', and represents the relative stretch of the metric. If two test masses are placed at a distance $L$, a passing GW would result in a oscillatory relative change of their distance $\Delta{L}/{L}\sim h$: the measurement of this relative change is at the basis of GW detection. The most promising GW detectors are laser interferometers. In these devices, the laser beam is split in the two arms of length $L$, and travels forth and back in the arms before being recombined. A passing GW causes a $\Delta{L}=L_2-L_1\sim hL$ that can be observed in the interference pattern of the recombined light. Earth based interferometers are limited in length (few Km) and cannot overcome the noise generated by gradients in the gravitational field of the Earth. They can therefore reach sensitivities of the order of $h\sim10^{-23}$ only in the kHz frequency range (See, for reference, the Adv. LIGO sensitivity curve in Figure \ref{fig1}). Stellar mass objects are therefore their main targets. To investigate lower frequencies, it is necessary to go to space. Here, the possibility of sending laser signals between spacecrafts orbiting at large separations from each other, together with the absence of gravity gradients, allows to probe the mHz frequency regime, with a strain sensitivity of the order of $h\sim10^{-20}$ (Figure \ref{fig1}). This is the realm of several classes of astrophysical objects including wide stellar binaries, EMRIs and MBH binaries. 

Precision timing of millisecond pulsars (MSPs) provide an alternative (though equivalent) way to detect GWs (see, e.g., \cite{saz78,hel83,jen05}). MSPs are the most stable natural clock in the Universe \cite{hobbs11}, and the time of arrival of their pulses can be currently predicted with a precision of the order of $\sim100ns$ (see, e.g., \cite{verbiest09}). If the pulsar and the Earth are in 'free-fall' in a spacetime perturbed by the passage of a GW, this will leave a characteristic fingerprint in the time of arrival of the radio pulses. As a matter of fact, the passing wave is 'changing the distance' between the pulsar and the Earth, which are now the 'two test masses' of our detector. The signal can be recovered by correlating the time of arrivals of an array of MSPs, i.e. a PTA, which can be (by analogy) thought as a 'galactic scale interferometer'. In this latter case, the sensitivity window is set by the duration of the timing experiment (typically several years) and by the sampling rate of the pulsars (typically once every few weeks), resulting in a range $\sim10^{-9}<f<10^{-6}$ Hz.

\section{Ground based interferometers}
\label{ligo}
The status of ground based interferometers and the expected Science delivered by their upcoming second generation is presented by Patrick Brady in his contribution to this proceeding series, here I just provide a general overview. Given their sensitivity curves (Figure \ref{fig1}), advanced ground-based interferometers are sensitive to stellar mass sources. Their main goal is the detection of inspiralling compact objects (NS-NS, NS-BH, BH-BH), and I report in table (\ref{tab1}) the expected detection rates, adapted from \cite{abadie10}. The large uncertainties in the figures are due to the poor observational constrains (we know only a handful of NS-NS systems which are going to coalesce in an Hubble time, \cite{kalogera04}), and to the many unknowns affecting theoretical population synthesis models (e.g., the physics of common envelope evolution \cite{dominik12}). GW observations will likely help to better understand binary evolution, possibly constraining the detailed physics of mass transfer and common envelope physics \cite{gerosa13}). NS deformation during the late inspiral of a NS-NS or NS-BH binary will inform us about the stiffness of the NS equation of state, probing the behavior of matter in extreme degenerate conditions (see, e.g., \cite{read09}). Coalescing NS-NS and NS-BH are also believed to be the engines powering short GRBs (see, e.g., \cite{koch93}), and coincident multimessenger (GW plus electromagnetic) detections will therefore provide the ultimate test for such hypothesis. Rotating NSs in the Galaxy are another potentially interesting source \cite{owen98}. Even though the strain is fairly small (see equation (\ref{strainspin})), many wave cycles will accumulate in a prolonged observation, resulting in an effective amplitude of the order of $h\sqrt{N}$, where $N=T/P$, being $P$ the period of the pulsar and $T$ the observation time (order of years). As shown in Figure \ref{fig1} Adv. LIGO might detect Galactic NSs spinning at kHz frequencies with $\epsilon\approx10^{-8}$. Collapsing and exploding objects (SNe, GRBs) are hard to model (See \cite{ott09} for a review), and different proposed scenarios result peculiar GW signatures roughly falling in the cyan area in Figure \ref{fig1}. Detection of GW bursts from SN explosions will therefore help to understand the physical mechanism driving the collapse and the subsequent catastrophic bounce. 
  
\begin{table}
\caption{Adv. LIGO detection rates, adapted from \cite{abadie10}. From the left to the right numbers are for pessimistic, realistic and optimistic rates (see \cite{abadie10} for details). The large range stems from the uncertainties in the complex physics involved in population synthesis models.}
\label{tab1}       
\begin{tabular}{l|lll}
\hline\noalign{\smallskip}
Source & $\dot{N}_{\rm low}$[yr$^{-1}$] & $\dot{N}_{\rm re}$[yr$^{-1}$] & $\dot{N}_{\rm high}$[yr$^{-1}$] \\
\noalign{\smallskip}\hline\noalign{\smallskip}
NS-NS & 0.4 & 40 & 400\\
NS-BH & 0.2 & 10 & 300\\
BH-BH & 0.4 & 20 & 1000\\
\noalign{\smallskip}\hline
\end{tabular}
\end{table}

\section{Space based interferometers}
\label{lisa}
\begin{figure}
  \includegraphics[width=5in]{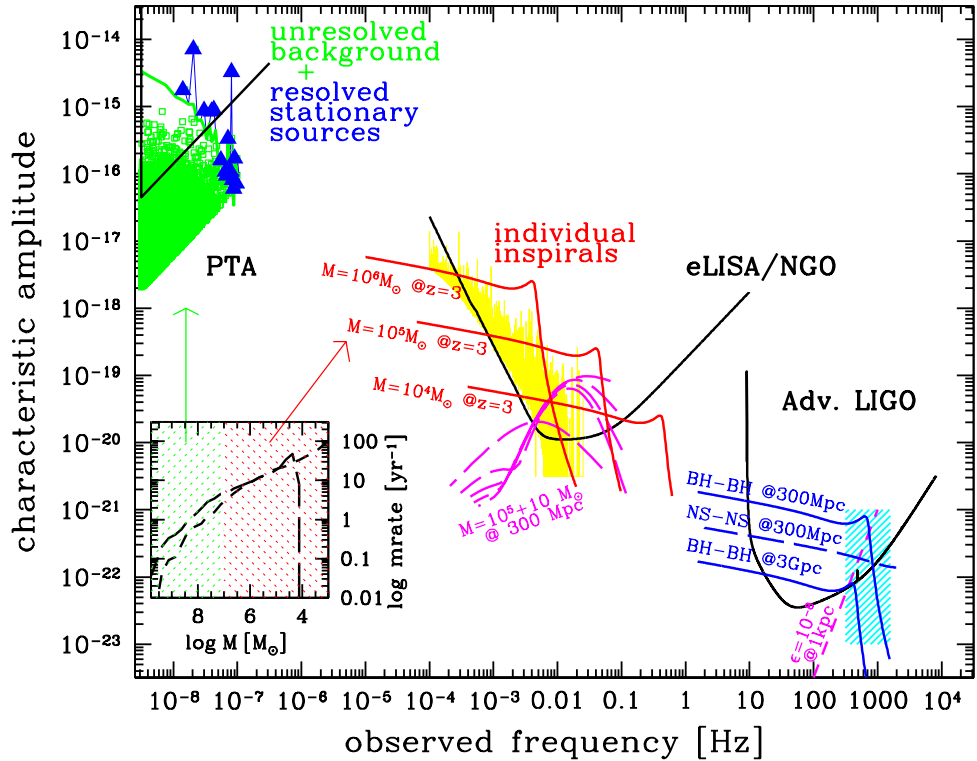}
\caption{Summary of the GW landscape discussed in the text. Black solid curves are detector sensitivities as labeled in the figure. For the PTA sensitivity curve we considered an SKA-like scenario in which 500 MSPs are monitored for 10 yrs at 50ns precision. In the Adv. In the LIGO band, the blue curves represent inspirals of compact stellar mass binaries \cite{thorne95}, the dashed magenta line marks the expected strain for rapidly spinning NSs \cite{owen98}, and the shaded cyan box identifies the area where burst signals from collapsing SNe are expected \cite{ott09}. In the eLISA/NGO band, the red curves represent merging MBH binaries at cosmological distances, the dashed-magenta line is a typical eccentric EMRI at $z=0.1$ (only the first 5 harmonics are shown) and the jagged yellow curve is the overall signal from a galactic population of WD-WD binaries. In the PTA band, the total signal (green solid line) is given by an incoherent superposition of individual MBH binaries (green dots); the brightest sources can be individually resolved (blue lines and triangles). The lower-left inset shows the inferred cosmological merger rate of MBH binaries as a function of mass for two specific models \cite{sesana11}. eLISA/NGO will probe the low mass part of the distribution (red shade) by detecting individual merging systems, whereas PTA will probe the high mass end (green shade) of the mass function by collecting signals from very massive systems in their inspiral phase.}
\label{fig1}      
\end{figure}

Space based interferometers like the proposed eLISA (which will be our default design in the following discussion) are sensitive in the mHz frequency band (Figure \ref{fig1}), where a rich variety of sources is expected. The enormous scientific payouts related to space based interferometers are far too many to be covered in a short review. The reader can only get a sense of it from the following discussion, a complete overview can be found in \cite{amaro12a,amaro12b}.

Space based interferometers are the only detectors for which we have 'verification sources' \cite{stroeer06}. These are known ultra-compact binaries emitting in the mHz range with high signal-to-noise ratio (SNR), and will serve as detector calibrators. Assuming the eLISA design, eight such binaries are known to date. Moreover eLISA will be sensitive to the whole population of wide WD-WD systems (order $10^8$ sources) populating the Galaxy. Most of those adds-up to the overall confusion foreground \cite{nelemans01,nissanke12}, but about 3000 of them will be individually resolved \cite{amaro12a,amaro12b}. Many resolvable binaries have periods between 5 and 10 minutes implying that they must have experienced at least one common-envelope phase; their statistics will therefore  provide critical tests of physical models of common-envelope binary evolution. Although rarer, eLISA will also detect a substantial population of short-period NS and BH binaries, determining their local merger rate. 

One of the most promising science goals of eLISA is the direct detection of MBHs, which appear to be a key component of galaxies. MBHs are ubiquitous in nearby massive galaxies (e.g., \cite{mago98}) and share a common evolutionary path with their hosts, undergoing repeated episodes of merger and accretion \cite{vhm03,malbon07}. Hierarchical models of MBH assembly predict that $10-100$ MBH mergers per year occur in the Universe (see, e.g., \cite{sesana07,sesana11}), the bulk of which involves systems with $10^4\, {\rm M_{\odot}}-10^6\,{\rm M_{\odot}}$ MBHs at $3<z<10$, falling in the sweetspot of the eLISA sensitivity curve. In fact, eLISA will be able to detect the GWs emitted by MBH binaries with total mass as small as $10^4\, {\rm M_{\odot}}$ and up to $10^8\,{\rm M_{\odot}}$, out to a redshift as remote as $z\sim 20$ with a SNR $> 10$. Redshifted masses will be measured to an unprecedented accuracy, up to the {0.1}--{1}{\%} level, whereas absolute errors in the spin determination are expected to be in the range {0.01}--{0.1}, allowing us to reconstruct their cosmic evolution \cite{sesana11}. eLISA observations hence have the potential of constraining the astrophysics of MBHs along their entire cosmic history, in a mass and redshift range inaccessible to conventional electromagnetic observations.  

On smaller scales, eLISA will also offer the deepest view of galactic nuclei, exploring the dynamics of stars in the space-time of a Kerr BH, by capturing the GWs emitted by stellar BHs orbiting the central MBH (EMRIs). A $10\, {\rm M_{\odot}}$ BH inspiralling into a $10^4\, {\rm M_{\odot}}-10^6\,{\rm M_{\odot}}$ MBH can be detected up to $z\simeq 0.5-0.7$ \cite{amaro12a,amaro12b}, allowing to explore a volume of several tens of Gpc$^3$ and to discover MBHs in dwarf galaxies that are still elusive to electromagnetic observations. Detailed measurement of the MBH mass and spin (up to one part in a thousand \cite{bc04}) will enable us to characterize the population of nuclear MBHs in an interval of masses where electromagnetic observations are poor, incomplete or even missing \cite{gair10}, providing information also on their spins. eLISA will also measure with equivalent precision the mass of the stellar BH in the EMRI event informing us about the mass spectrum of stellar BHs, which is basically unknown.

High SNR measurement of the GW waveforms will allow to test GR in the highly dynamical strong field regime of MBHs, including no-hair theorem tests, constrains on a putative graviton mass and on alternative theory of gravity in general. More details about testing gravity with GW observations are given by Emanuele Berti in his contribution to this proceedings series \cite{berti13}.

\section{Pulsar timing arrays}
\label{pta}
By exploiting the characteristic fingerprint left by passing GWs in the time of arrival of the radio pulses propagating from the pulsar to the receiver on Earth \cite{saz78,hel83,jen05}, PTAs  are sensitive to the collective signal coming from the inspiralling population of supermassive low redshift binaries \cite{sesana08,sesana09} in the nHz frequency band. The overall expected characteristic strain $h_c$ can be written as
\begin{equation}
h_c^2(f) =\int_0^{\infty} 
dz\int_0^{\infty}d{\cal M}\, \frac{d^3N}{dzd{\cal M} d{\rm ln}f_r}\,
h^2(f_r),
\label{hch2}
\end{equation}
where $d^3N/dzd{\cal M} d{\rm ln}f_r$, is the comoving number of binaries emitting in a given logarithmic frequency interval with chirp mass and redshift in the range $[{\cal M},{\cal M}+d{\cal M}]$ and $[z, z+dz]$, respectively; and $h(f_r)$ is the sky and polarization averaged version of the strain amplitude given in equation (\ref{eqstrain}), and is given by ~\cite{tho87}:
\begin{equation}
h={8\pi^{2/3}\over 10^{1/2}}{{\cal M}^{5/3}\over D_L(z)}f_r^{2/3}\,,
\label{eqthorne}
\end{equation}
where $D_L(z)$ is the luminosity distance to the source. The GW spectrum has a characteristic shape $h_c=A(f/1{\rm yr}^{-1})^{-2/3}$ (green solid line in Figure \ref{fig1}), where $A$ is the signal normalization at $f=1{\rm yr}^{-1}$, which depends on the details of the MBH binary population only. Therefore, GW detection will provide a strong test for the effectiveness of the merger process of galaxies and their MBHs at low redshift, and will help constraining the high mass end of the MBH mass function. Recent works \cite{ravi12,mcwilliams12,sesana12} set a plausible range $3\times10^{-16}<A<4\times10^{-15}$, the upper limit being already in tension with current PTA measurement \cite{vh11,demorest12}. With few more years of  observation, PTAs have a concrete chance of detecting this signal. Several millions of sources contribute to it, however, the bulk of the strain comes from few hundred sources only (see, e.g., \cite{kocsis11}) generating a signal at a 1ns level or higher, considered the ultimate goal for the SKA. Therefore, the signal is far from being a Gaussian isotropic background \cite{ravi12}; a handful of sources dominate the strain budget, and might be individually identified. Interestingly, such systems are far from coalescence, and they can still retain much of their original eccentricity against GW circularization \cite{sesana10,preto11,roedig11}. Eccentricity measurement of individually resolved sources may help in constraining the evolution of MBH binaries, testing our current models of their dynamical evolution in star and/or gas dominated environments. For bright enough sources (SNR$\approx10$) sky location within few tens to few deg$^2$ is possible \cite{sesanavecchio10,ellis12,petiteau13} (and even sub deg$^2$ determination, under some specific conditions \cite{kj11}). Even though this is a large chunk of the sky, these systems are extremely massive and at relatively low redshift ($z<0.5$), making any putative electromagnetic signature of their presence (emission periodicity related to the binary orbital period, peculiar emission spectra, peculiar K$\alpha$ line profiles, etc.) detectable \cite{sesana11b,tanaka11}.  

\section{Conclusions}
\label{concl}
Gravitational wave astronomy has never been so active as in the last few years. The prospect of the first detection from Advanced ground based interferometers has triggered a wave of excitement in the astrophysical community, pushing the concept of multimessenger astronomy and the development of pipelines that combine GW and electromagnetic interactions to unveil the nature of short gamma ray bursts and supernovae. A worldwide effort of detecting low frequency waves (in the nHz regime) by timing ultra-precise millisecond pulsars is running in parallel. The enormous advanced in the quality of the receivers and in the data reduction techniques is making the goal of achieving sub-100ns timing precision on a bunch of pulsars possible. Current array sensitivities are a factor of a few away from theoretical prediction of the nHz signal expected from cosmological supermassive black holes, making detection within this decade plausible. Looking a step ahead in the future, space based interferometer such as LISA or eLISA/NGO will cover the mHz regime, providing an almost full coverage of the gravitational wave spectrum from the nHz to the kHz. The silent side of the Universe will soon get loud.




\begin{thebibliography}{}
%
%

\bibitem{harry10}
Harry G. M., and the LIGO Scientific Collaboration, 2010, Classical and Quantum Gravity, 27, 4006
\bibitem{acernese06}
Acernese F., et al., 2006, Classical and Quantum Gravity, 23, 63
\bibitem{somiya12}
Somiya K., et al., 2012, Classical and Quantum Gravity, 29, 4007
\bibitem{punturo10}
Punturo M., et al., 2010, Classical and Quantum Gravity, 27, 4002
\bibitem{bender98}
Bender P.-L. et al., 1998, LISA Pre-Phase A Report, Second Edition, MPQ
233
\bibitem{amaro12a}
Amaro-Seoane P., et al., 2012, Arxiv e-prints 1201.3621
\bibitem{amaro12b}
Amaro-Seoane P., et al., 2012, Classical and Quantum Gravity, 29, 4016
\bibitem{ferdman10}
Ferdman R. D., et al., 2010, Classical and Quantum Gravity, 27,
084014
\bibitem{man12}
Manchester R. N., et al., 2012, ArXiv e-prints 1210.6130
\bibitem{jenet09}
Jenet F. A., et al., 2009, ArXiv e-prints 0909.1058
\bibitem{hobbs10}
Hobbs G., et al., 2010, Classical and Quantum Gravity, 27, 084013
\bibitem{aha13}
Aharonian F., et al., 2013, Arxiv e-prints 1301.4124
\bibitem{binetruy13}
Binetruy P., Boh\`e A., Caprini C. \&  Dufaux J.-F., 2012, JCAP, 6, 27
\bibitem{thorne95}
Thorne, K. S., Particle and Nuclear Astrophysics and Cosmology in the Next Millenium, Proceedings of the 1994 Snowmass Summer Study held 29 June - 14 July, 1994. Edited by E.W. Kolb and R.D. Peccei. Singapore: World Scientific, 1995., p.160
\bibitem{hughes03}
Hughes S. A., 2003, Annals of Physics, Volume 303, Issue 1, p. 142-178
\bibitem{einstein16}
A. Einstein, 1916, Koeniglich Preu\S ische Akademie der Wissenschaften Berlin,
Sitzungsberichte, 688.
\bibitem{damour83}
Damour T., 1983, Phys. Rev. Lett., 51, 1019.
\bibitem{taylor89}
Taylor J. H. \& Weisberg J. M., 1989, Astrophys. J., 345, 434
\bibitem{ott09}
Ott C. D., 2009, Classical and Quantum Gravity, 26, 3001
\bibitem{saz78}
 Sazhin M. V., 1978, Soviet Astron., 22, 36
\bibitem{hel83}
 Hellings R. W. \& Downs G. S., 1983, Astrophys. J., 265, 39
\bibitem{jen05}
Jenet F. A., Hobbs G. B., Lee K. J. \& Manchester R. N., 2005, Astrophys. J., 625, 123
\bibitem{hobbs11}
Hobbs G., et al., 2011, MNRAS, 427, 2780
\bibitem{verbiest09}
Verbiest J. P. W., et al., 2009, MNRAS, 400, 951
\bibitem{owen98}
Owen B. J., Lindblom L., Cutler C., Schutz B. F., Vecchio A. \& Andersson N., 1998, Phys. Rev. D, 58, 4020
\bibitem{sesana11}
Sesana A., Gair J., Berti E. \& Volonteri M., 2011, Phys. Rev. D, 83, 4036
\bibitem{abadie10}
Abadie J., et al., 2010, Classical and Quantum Gravity, 27, 3001
\bibitem{kalogera04}
Kalogera V., et al., 2004, Astrophys. J. Lett., 601, 179
\bibitem{dominik12}
Dominik M., et al., 2012. Astrophys. J., 759, 52
\bibitem{gerosa13}
Gerosa D., Kesden M., Berti E., O'Shaughnessy R. \& Sperhake U., 2013, ArXiv e-prints 1302.4442
\bibitem{read09}
Read J., et al., 2009, Phys. Rev. D, 79, 4033
\bibitem{koch93}
Kochanek C. S. \& Piran T., 1993, Astrophys. J. Lett., 417, 17
\bibitem{stroeer06}
Stroeer A. \& Vecchio A., 2006,  Classical and Quantum Gravity, 23, 809
\bibitem{nelemans01}
Nelemans G., Yungelson L. R., Portegies Zwart S. F., 2001, A\&A, 375, 890
\bibitem{nissanke12}
Nissanke S., Vallisneri M., Nelemans G. \& Prince T. A., 2012, Astrophys. J., 758, 131
\bibitem{mago98}
Magorrian J. et al., 1998, AJ, 115, 2285
\bibitem{vhm03}
Volonteri M., Haardt F. \& Madau P., 2003, Astrophys. J., 582, 599 
\bibitem{malbon07}
Malbon, R. K., Baugh, C. M., Frenk, C. S., \& Lacey, C. G. 2007, MNRAS, 382, 1394
\bibitem{sesana07}
Sesana A., Volonteri M. \& Haardt F., 2007, MNRAS, 377, 1711
\bibitem{bc04}
Barack L. \& Cutler C., 2004, Phys. Rev. D., 69, 2005
\bibitem{gair10}
Gair J. R., Tang C. \& Volonteri M., 2010, Phys. Rev. D, 81, 4014
\bibitem{berti13}
Berti E., 2013, ArXiv e-prints 1302.5702
\bibitem{sesana08} 
Sesana A., Vecchio A. \& Colacino C. N., 2008, MNRAS, 390, 192
\bibitem{sesana09} 
Sesana A., Vecchio A. \& Volonteri M., 2009, MNRAS, 394, 2255
\bibitem{tho87} 
Thorne K. S., 1987, Gravitational radiation.. pp 330–458
\bibitem{ravi12}
Ravi V., et al., 2012, Astrophys. J., 761, 84
\bibitem{mcwilliams12}
McWilliams S. T., Ostriker J. P., Pretorius F., 2012, ArXiv e-prints 1211.4590
\bibitem{sesana12}
Sesana A., 2012, MNRAS, in press, ArXiv e-prints 1211.5375
\bibitem{vh11}
van Haasteren R., et al., 2011, MNRAS, 414, 3117
\bibitem{demorest12}
Demorest P. B., et al., 2013, Astrophys. J., 762, 94
\bibitem{kocsis11} 
Kocsis B. \& Sesana A., 2011, MNRAS, 411, 1467
\bibitem{sesana10} 
Sesana A., 2010, ApJ, 719, 851
\bibitem{preto11}
Preto M., Berentzen I., Berczik P. \& Spurzem R., 2011, Astrophys. J., 732, 26
\bibitem{roedig11} 
Roedig C., Dotti M., Sesana A., Cuadra J. \& Colpi M., 2011, MNRAS, 415, 3033
\bibitem{sesanavecchio10}
Sesana A. \& Vecchio A., 2010, Phys. Rev. D, 81, 4008 
\bibitem{ellis12}
Ellis J. A., Siemens X. \& Creighton J. D. E., 2012, Astrophys. J., 756, 175
\bibitem{petiteau13}
Petiteau A., Babak S., Sesana A. \& de Araujo M., 2013, Phys. Rev. D, in press, ArXiv e-prints 1210.2396
\bibitem{kj11}
Lee K. J., et al., 2011, MNRAS, 414, 3251
\bibitem{sesana11b} 
Sesana A., Roedig C., Reynolds M.T. \& Dotti M., 2012, MNRAS, 420, 860
\bibitem{tanaka11} 
Tanaka T., Menou K. \& Haiman Z., 2012, MNRAS, 420, 705


\end{thebibliography}


\end{document}